\documentclass[%
 reprint,
 amsmath,amssymb,
 aps,
 pra
]{revtex4-2}

\usepackage{graphicx}%
\usepackage{dcolumn}%
\usepackage{bm}%
\usepackage{booktabs}
\usepackage{hyperref}%
\usepackage{xcolor}
\usepackage{soul}

\begin{document}

\preprint{APS/123-QED}

\title{ Boundaries of universality of thermal collisions for atom - atom scattering}%

\author{Xuyang Guo}
 \affiliation{%
 Department of Chemistry, University of British Columbia,
Vancouver, B.C. V6T 1Z1, Canada%
 }%

 \author{Kirk W. Madison}%
 \affiliation{%
 Department of Physics and Astronomy, University of British Columbia, Vancouver, B.C. V6T 1Z1, Canada%
 }%

\author{James L. Booth}
\affiliation{
    Physics Department, British Columbia Institute of Technology,
   Burnaby, B.C. V5G 3H2, Canada%
}%

\author{Roman V. Krems}
\email[Corresponding author: ]{rkrems@chem.ubc.ca}
\affiliation{Department of Chemistry \& Stewart Blusson Quantum Matter Institute, University of British Columbia,
Vancouver, B.C. V6T 1Z1, Canada 
}%

\date{\today}%

\begin{abstract}

Thermal rate coefficients for some atomic collisions have been observed to be remarkably independent of the details of interatomic interactions at short range.  
This makes these rate coefficients universal functions of the long-range interaction parameters and masses, which was previously exploited to develop a self-defining atomic sensor for ambient pressure.
Here, we employ rigorous quantum scattering calculations to 
examine the response of thermally averaged rate coefficients for atom - atom collisions to changes in the interaction potentials.
We perform a comprehensive analysis of the universality, and the boundaries thereof, by treating the quantum scattering observables as probabilistic predictions determined by a distribution of interaction potentials.
We show that there is a characteristic change of the resulting distributions of rate coefficients, separating light, few-electron atoms and heavy, polarizable atoms.
We produce diagrams that illustrate the boundaries of the thermal collision universality at different temperatures and provide guidance for future experiments seeking to exploit the universality.

\end{abstract}

\maketitle

\section{Introduction}

Within the Born-Oppenheimer approximation, the scattering cross sections for collisions between atoms are assumed to be determined by the adiabatic interaction potentials, which parametrize the Schr\"{o}dinger equation for the collision complex. 
The interaction potentials are generally obtained by electronic structure calculations with approximations that introduce errors limiting the accuracy of quantum scattering calculations. The effect of these errors on numerical predictions of quantum dynamics 
observables, such as collision cross sections or scattering rates, for heavy multi-electron atoms is not well understood. At the same time, numerical solutions of quantum scattering equations are used for astrophysical models, for benchmarking experimental measurements and 
as calibration parameters for various experimental techniques. For example, Refs.~\cite{madison2012cold,booth54METHODDEVICE2014,boothUniversalityQuantumDiffractive2019,shenRealizationUniversalQuantum2020,shenRefiningColdAtom2021,stewartMeasurementRbRbVan2022,barkerAccurateMeasurementLoss2023,shenCrosscalibrationAtomicPressure2023,frielingCrosscalibrationQuantumAtomic2024} detail the use of laser-cooled atoms to measure the density of ambient gases in a vacuum by observing collision-induced heating and loss rate of trapped sensor atoms. The relation between the observed trap loss and density is through a collision cross section, which is assumed to be known from an independent source, often a theoretical calculation.  For vacuum metrology based on trapped sensor atoms~\cite{boothUniversalityQuantumDiffractive2019}, the current goal is to reduce the uncertainty of observables to less than 1\%. 
However, it is not clear if electronic structure calculations can produce interaction potentials for heavy, multi-electron atoms and molecules with errors small enough to meet these accuracy requirements.  
For such applications, it is necessary (i) to be able to quantify the errors of the numerical predictions of quantum dynamics observables; and (ii) design the experiments within the range of parameters and with atomic species that reduce the effect of errors of interatomic interaction potentials on quantum scattering calculations.

Recent experimental work on measuring pressure of H$_2$, He, Ar, CO$_2$, Kr and Xe  gases with trapped $^{87}$Rb atoms~\cite{shenRefiningColdAtom2021,stewartMeasurementRbRbVan2022,barkerAccurateMeasurementLoss2023,shenRealizationUniversalQuantum2020}, accompanied by scattering calculations with model Lennard-Jones potentials~\cite{shenRealizationUniversalQuantum2020,boothRevisingUniversalityHypothesis2024,boothUniversalityQuantumDiffractive2019} and selected rigorous calculations for atom - atom and atom - molecule collisions~\cite{boothUniversalityQuantumDiffractive2019,makridesElasticRateCoefficients2019,makridesCollisionsRoomtemperatureHelium2020,makridesErratumCollisionsRoomtemperature2022,makridesErratumElasticRate2022,klosElasticGlancingangleRate2023,shenCrosscalibrationAtomicPressure2023, herpergerDeviationQuantumDiffraction2023}, illustrated the invariance of room-temperature thermal collision rate coefficients to variations of the short-range part of the underlying atom - atom and atom - molecule potentials. The rates for trap loss induced by collisions with atoms in a thermal background gas were found to be universal functions of the trap depth, the long-range interaction parameters, and the mass of the trapped sensor atom~\cite{childMolecularCollisionTheory2014,boothUniversalityQuantumDiffractive2019,boothRevisingUniversalityHypothesis2024,shenRealizationUniversalQuantum2020,shenRefiningColdAtom2021}. At the same time, it was observed that certain background species, most notably light few-electron atoms and molecules, exhibited significant deviations from the predicted universal behavior. 
Furthermore, experiments with two trapped species, $^7$Li and $^{87}$Rb, performed in two different laboratories~\cite{barkerAccurateMeasurementLoss2023,frielingCrosscalibrationQuantumAtomic2024} produced systematically different results for the $^7$Li/$^{87}$Rb ratios of thermal rate coefficients. 
K{\l}os and Tiesinga~\cite{klosElasticGlancingangleRate2023} performed rigorous scattering calculations, including an estimate of calculation uncertainties, to calibrate the measurements in Ref.~\cite{barkerAccurateMeasurementLoss2023}. 
It was found that the ratios of rate coefficients for collisions with He,  Ne, Kr, and Xe agree with the experimental results, but disagree for Ar. 
However, 
the most recent cross-calibration experiment combining two traps loaded with $^7$Li and $^{87}$Rb~\cite{frielingCrosscalibrationQuantumAtomic2024} to measure the $^7$Li/$^{87}$Rb ratios of the rate coefficients for collisions with He, Ne, Ar, Kr, and Xe,
revealed systematic differences of up to 3\% from the theoretically predicted ratios~\cite{klosElasticGlancingangleRate2023}. For Ar, Kr and Xe, these differences exceeded the stated uncertainties of the theoretical calculations.    

The collision universality can liberate measurement analyses and calibration from the uncertainties of electronic structure calculations. It is therefore important to 
identify the origin and quantify the boundaries of universality. To achieve this, 
we perform a comprehensive analysis of the universality of rate coefficients for atom - atom elastic scattering. This work is based on rigorous quantum scattering calculations employing the most accurate available interaction potentials as a reference.  
We illustrate the collision universality by treating the quantum scattering observables as probabilistic predictions determined by a distribution of interaction potentials~\cite{moritaUniversalProbabilityDistributions2019}. 
We show that for heavy background gas atoms (Ar, Kr, Xe), the distributions of rate coefficients are highly concentrated, yielding consistent collision rate coefficients over a wide range of interaction potentials. At the same time, the variance of the distributions for light background gas atoms is shown to be very large. This characteristic change of the rate coefficient distributions marks the transition from the universal scattering regime to thermal scattering that is sensitive to the short-range part of the interaction potentials. 
We explore the variance of the resulting distributions of observables as a function of long-range interaction parameters, the mass of the background gas atoms, and the temperature. This yields diagrams that illustrate the limitations of the thermal collision universality at different temperatures and provide guidance for future experiments seeking to exploit the universality. Finally, we examine in detail the discrepancy of the experimental measurement~\cite{frielingCrosscalibrationQuantumAtomic2024} from the theoretical prediction~\cite{klosElasticGlancingangleRate2023} and show that a 2-4\% change in the long-range interaction parameters can account for the differences between the experimental measurements and the results of the quantum scattering calculations.

\section{Calculation details}

\subsection{Scattering calculations}
\label{sec:qsc}
We compute and analyze rate coefficients for collisions of trapped atoms ($^7$Li and $^{87}$Rb) with atoms of mass $m$ in an ambient thermal gas (background gas BG) with temperature $T$
by integrating the energy dependence of the cross sections $\sigma(E)$ with the Maxwell-Boltzmann distribution as follows: 
\begin{equation}
    \kappa(T)=  \sqrt{\frac{8 k_{\rm B} T}{\pi m}} \frac{1}{\left ( k_{\rm B} T \right )^2}
    \int_0^{\infty}  E \sigma(\tilde{E}) \exp{\left (-\frac{E}{k_{\rm B}T} \right )} \mathrm{d} E,
        \label{eq:cal_rate}
    \end{equation} 
    where $k_{\rm B}$ is the Boltzmann constant, $E$ is the kinetic energy of the incident particle, {$\tilde{E}=\mu E/m$ is the collision energy. Assuming the incident particle velocity is the same as the relative velocity in the center-of-momentum (COM) frame is exact in the limit where the trapped sensor particle is stationary. This approximation is justified here because the energy of the sensor atoms in the trap is approximately $10^6$ times smaller than the energy of the background gas particles~\cite{Avinash2024}.
    }
We consider collisions of $^7$Li and $^{87}$Rb with closed-shell atoms He, Ne, Ar, Kr, or Xe.  Such collisions do not permit inelastic scattering at collision energies relevant for this work.

The collision cross sections $\sigma$ are obtained from the numerical solutions of the time-independent Schr\"{o}dinger equation, which, upon partial wave expansion of the full wave function, can be written as
\begin{equation}
    \left[\frac{\mathrm{d}^2}{\mathrm{~d} R^2} + k^2 - \frac{l(l+1)}{R^2}\right] F_{l}(R) =  - \frac{2 \mu}{\hbar^2} V(R) F_{l}(R),
    \label{eq:itse_cc}
\end{equation} 
where $R$ is the distance between the centers of mass of the colliding atoms, $l$ is the rotational angular momentum of the collision complex, $V(R)$ is the adiabatic interaction potential, 
\begin{eqnarray}
k^2 =  \frac{2 \mu {\tilde{E}}}{\hbar^2},
\end{eqnarray}
 $\mu$ is the reduced mass of the collision complex. 
We use the log-derivative method~\cite{manolopoulosImprovedLogDerivative1986} to compute {$Y(R)=\frac{1}{F_l(R)} \frac{\mathrm{d} F_l(R)}{\mathrm{d} R}$} and match the numerical solutions to the scattering boundary conditions, yielding scattering phase shifts $\eta_l$. The scattering cross section is then computed as
\begin{equation}
    \sigma({\tilde{E}})=\frac{4\pi }{k^2} \sum_{l=0}^{\infty} (2 l+1)\sin ^2 \eta_l.
    \label{partial-wave}
\end{equation} 
We include enough terms in the partial wave sum (\ref{partial-wave}) to ensure convergence of the cross section to $\leq 0.1\%$ and extend the range of collision energies in Eq.~(\ref{eq:cal_rate}) to ensure convergence of the thermal rate coefficients to $\leq 0.25\%$. {The number of partial waves required for convergence depends on the reduced mass of the system and the collision energy $\tilde{E}$. We use the values of $l$ up to 200 for the Li-He system (lowest reduced mass) and up to 1200 for the Rb-Xe system (largest reduced mass considered in this work).}

For analysis purposes, we also compute the rate coefficients $\kappa (T)$ using the scattering phase shifts given by the semiclassical Jeffreys-Born approximation~\cite{boothRevisingUniversalityHypothesis2024}
\begin{eqnarray}
    \eta_l(k)=&& \frac{3\pi}{16}\left (\frac{\mu C_6}{\hbar^2} \right )\frac{k^4}{(l+\frac{1}{2})^5} + \frac{5\pi}{32}\left (\frac{\mu C_8}{\hbar^2}\right )\frac{k^6}{(l+\frac{1}{2})^7} \nonumber\\ 
    && + \frac{35\pi}{256}\left (\frac{\mu C_{10}}{\hbar^2} \right )\frac{k^8}{(l+\frac{1}{2})^9},
    \label{eq:JBJB}
\end{eqnarray} 
where $C_n$ are the coefficients of the long-range expansion of the interaction potential
\begin{eqnarray}
\lim_{R \rightarrow \infty} V(R) = - \frac{C_6}{R^6} - \frac{C_8}{R^8} - \frac{C_{10}}{R^{10}}.
\label{LR}
\end{eqnarray}
The rate coefficients thus computed are determined entirely by the long-range part of the interaction potential.

\subsection{Atom-atom interaction potentials}
The goal of the present work is to examine the response of the thermal collision rate coefficients (\ref{eq:cal_rate})  to variations of the interaction potentials $V(R)$ at short range $R$. To achieve this, we model $V(R)$
by
the Morse long-range (MLR) functions~\cite{medvedevInitioInteratomicPotentials2018}:
\begin{equation}
   V(R)=D_{\mathrm{e}}\left(1-\frac{u_{\mathrm{LR}}(R)}{u_{\mathrm{LR}}\left(R_{\mathrm{e}}\right)} \exp \left[-\beta(R) \cdot y_p^{\mathrm{eq}}(R)\right]\right)^2,
    \label{eq:mlr}
\end{equation} 
where $D_e$ is the potential energy at the interatomic equilibrium distance $R_e$ and $\beta(R)$ is a polynomial function
\begin{equation}
    \beta(R)=y_p^{\mathrm{ref}} \beta_{\infty}+\left[1-y_p^{\mathrm{ref}}\right] \sum_{n=0}^N \beta_i\left[y_q^{\mathrm{ref}}\right]^n
    \label{eq:ref_dist}
    \end{equation} of the reduced coordinates {$y_{j}^{\mathrm{ref}}(R)$ defined as
    \begin{equation}
    y_{j}^{\mathrm{ref}}(R)=\frac{R^{j}-R_{\mathrm{ref}}^{j}}{R^{j}+R_{\mathrm{ref}}^{j}}
    \end{equation} with $j \in\{p, q\}$ and} $R_{\mathrm{ref}}$ representing a reference distance~\cite{leroyAccurateAnalyticPotentials2009}. 
Following Ref.~\cite{medvedevInitioInteratomicPotentials2018}, 
we fix $p=4$ and $q=5$. 

The radial variable $y_p^{\mathrm{eq}}$ is given by Eq.~(\ref{eq:ref_dist}) 
with $R_{\mathrm{ref}}=R_e$. The constraint parameter $\beta_{\infty}$ is fixed to $\beta_{\infty} = \ln \left\{2 D_{\mathrm{e}} / u_{\mathrm{LR}}\left(R_{\mathrm{e}}\right)\right\}$. 
Finally, the long-range part of the potential $u_{\mathrm{LR}}$ is computed as
\begin{equation}
    u_{\mathrm{LR}}(R)= D_6 \frac{C_6}{R^6} + D_8 \frac{C_8}{R^8}+ D_{10} \frac{C_{10}}{R^{10}},
    \end{equation} where $D_n$ are the damping functions given by the generalized Douketis function~\cite{douketisIntermolecularForcesHybrid1982a,leroyLongrangeDampingFunctions2011}:
    \begin{equation}
        D_n(r,s)=\left(1-\exp \left(-\frac{b(s)(\rho R)}{n}-\frac{c(s)(\rho R)^2}{n^{1 / 2}}\right)\right)^{n+s},
        \end{equation} 
        where $b(s)$ and $c(s)$ are system-independent parameters determined by the choice of $s$. 
        In this work, we use $s=-1$, $b = 3.3$, and $c=0.423$, as in  Ref.~\cite{medvedevInitioInteratomicPotentials2018}. The dimensionless parameter \(\rho = \frac{2 \rho_{\mathrm{M}} \rho_{\mathrm{BG}}}{\left( \rho_{\mathrm{M}} + \rho_{\mathrm{BG}} \right)}\)~\cite{douketisIntermolecularForcesHybrid1982a} is calculated as the harmonic mean of the atomic counterparts \(\rho_i = \left( \frac{\mathrm{IP}_{\mathrm{i}}}{\mathrm{IP}_{\mathrm{H}}} \right)^{\frac{2}{3}}\), where \(\mathrm{IP}_{\mathrm{i}}\) and \(\mathrm{IP}_{\mathrm{H}}\) represent the ionization potentials of the respective atom and hydrogen. 
  For Rb-BG interactions (BG = He, Ne, Ar, Kr, Xe), we use the MLR parameters from Ref.~\cite{medvedevInitioInteratomicPotentials2018}. For Li-BG interactions, we obtain MLR parameters by fitting the {\it ab initio} interaction potentials from Ref.~\cite{klosElasticGlancingangleRate2023}, with the fitting root mean squared error $ < 1~\mathrm{cm}^{-1}$. {The parameters for Li-BG interactions are given in the supplemental material~\cite{supplemental}.}

    \begin{table}[b]
      \caption{\label{tab:q_sc_dev}
      Collision rate coefficients (in units of $\times~10^{-15}~\mathrm{m^3/s}$) calculated at $T = 294.15$ K for different collision systems from quantum (Q) and semi-classical (SC) calculations. }
      \begin{ruledtabular}
      \begin{tabular}{ccccc}
      Sensor & Gas & $\kappa_{\rm Q}$ & $\kappa_{\rm SC}$ & $D$ of Eq.~(\ref{eq:def_uni}) in \% \\ 
      \hline
      Rb & He & 2.444 & 3.077 & -20.57 \\ 
      Rb & Ne & 1.985 & 2.288 & -13.22 \\ 
      Rb & Ar & 3.045 & 3.012 & 1.07 \\ 
      Rb & Kr & 2.794 & 2.780 & 0.51 \\ 
      Rb & Xe & 2.889 & 2.890 & -0.03 \\ 
      Li & He & 1.647 & 2.295 & -28.23 \\ 
      Li & Ne & 1.559 & 1.705 & -8.53 \\ 
      Li & Ar & 2.332 & 2.293 & 1.69 \\ 
      Li & Kr & 2.146 & 2.113 & 1.52 \\ 
      Li & Xe & 2.237 & 2.198 & 1.79 \\ 
      \end{tabular}
      \end{ruledtabular}
      \end{table}

      \begin{table}[b]
        \caption{ \label{tab:rate_distrib} Mean, and relative standard deviation (RSD) in \% of the distributions of $\kappa$ for the ranges of $R_e \in[0.25,7.25]\,\text{\AA}$ and $ D_e \in [1,110]~\mathrm{cm}^{-1}$, excluding unphysical potentials. Mean are in the units of $\times~10^{-15}~\mathrm{m^3/s}$. RSD$^*$ denote the relative standard deviation of the distribution of $\kappa$ for the ranges of $R_e \in [0.8 \times R_{e,0},1.2 \times R_{e,0}]$ and $D_e \in [0.8 \times D_{e,0},1.2 \times D_{e,0}]$, where $R_{e,0}$ and $D_{e,0}$ are the well depth and the equilibrium distance from the  MLR fits (\ref{eq:mlr}) of the corresponding {\it ab initio} potentials. }
          \begin{ruledtabular}
            \begin{tabular}{cccccc}
              Sensor & Gas & Mean & RSD \% & RSD$^*$\% & Universality \\ \hline
              Rb & He & 3.573 & 27.1 & 41.4 & Non-universal \\ 
              Rb & Ne & 2.418 & 13.7 & 9.1 & Non-universal \\ 
              Rb & Ar & 3.043 & 5.3 & 1.2 & Universal \\ 
              Rb & Kr & 2.787 & 3.2 & 0.5 & Universal \\ 
              Rb & Xe & 2.881 & 2.8 & 0.2 & Universal \\ 
              Li & He & 2.851 & 44.0 & 35.4 & Non-universal \\ 
              Li & Ne & 1.939 & 20.9 & 8.2 & Non-universal \\ 
              Li & Ar & 2.350 & 8.3 & 1.5 & Universal \\ 
              Li & Kr & 2.162 & 4.1 & 0.5 & Universal \\ 
              Li & Xe & 2.243 & 2.5 & 0.3 & Universal \\ 
            \end{tabular}
              \end{ruledtabular}
         
      \end{table}

\section{Results}

    \subsection{Universality of thermal collisions}
    \label{sec:ill_uni}
 The collision universality introduced in Refs.~\cite{childMolecularCollisionTheory2014,boothUniversalityQuantumDiffractive2019,boothRevisingUniversalityHypothesis2024,shenRealizationUniversalQuantum2020,shenRefiningColdAtom2021} manifests itself in the invariance of thermal collision rate coefficients (\ref{eq:cal_rate}) to changes in the short-range part of the adiabatic interaction potentials. 
 The rate coefficients are thus universal functions of the long-range interaction parameters and the masses of the collision partners. 
  This universality is particularly important for vacuum metrology as it ensures that total scattering rate coefficients—and therefore the inferred gas density or pressure—remain unaffected by the uncertainty in the short-range part of the interaction potentials~\cite{childMolecularCollisionTheory2014,boothUniversalityQuantumDiffractive2019,boothRevisingUniversalityHypothesis2024,shenRealizationUniversalQuantum2020,shenRefiningColdAtom2021}. 
  In this section, we examine in detail the response of thermal rate coefficients to variations of interaction potentials for a range of different collision systems in order to elucidate the origin of the universal behavior of rate coefficients observed in Refs.~\cite{childMolecularCollisionTheory2014,boothUniversalityQuantumDiffractive2019,boothRevisingUniversalityHypothesis2024,shenRealizationUniversalQuantum2020,shenRefiningColdAtom2021}.   
          \begin{figure}[b]
                \includegraphics[width=0.45\textwidth]{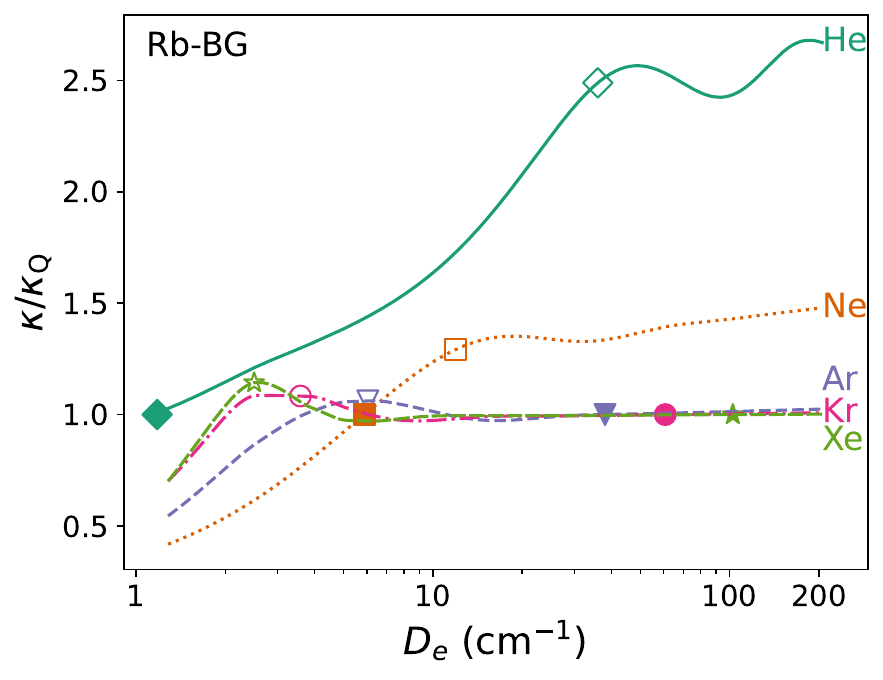}
                \includegraphics[width=0.45\textwidth]{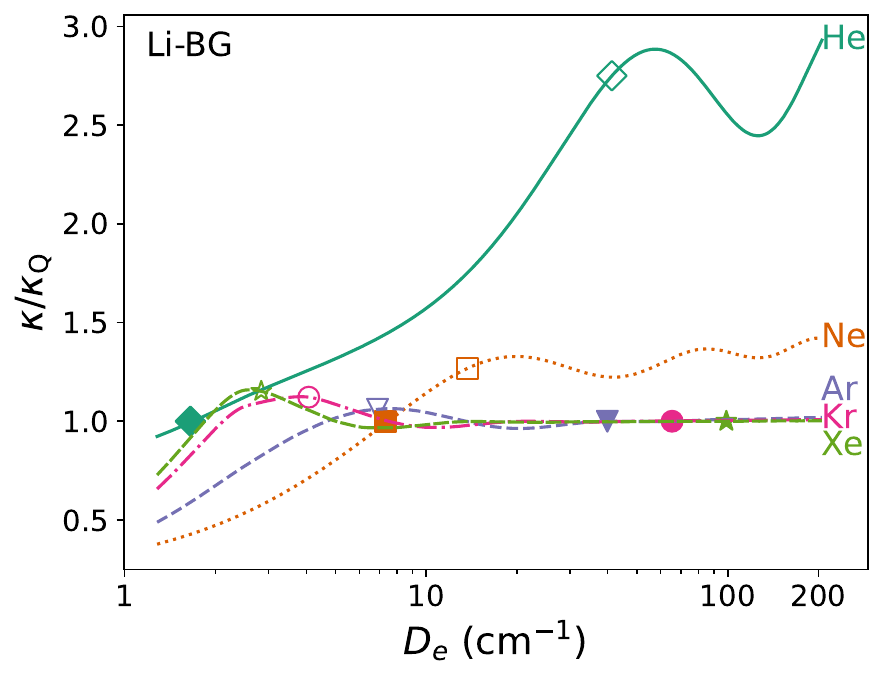}
            \caption{Dependence of room-temperature rate coefficient ratios $\frac{\kappa}{\kappa_Q}$ for Rb -- BG (upper panel) and Li -- BG (lower panel) scattering on the well depth $D_e$ of the modified interaction potentials. The {solid symbols} show the values of $D_e$ for the best fit of the {\it ab initio} interaction potentials for 
He (cyan-blue, diamonds), Ne (orange, squares), Ar (purple, triangles), Kr (pink, circles), and Xe (green, stars). The rate coefficients are normalized by the magnitude of the rate coefficient at these values of $D_e$. 
The open symbols indicate the values of $D_e$ corresponding to $ m (0.48 R_0 D_e/\hbar)^2/2 = k_{\rm B}T$~\cite{boothRevisingUniversalityHypothesis2024}.
Rate coefficients for interactions potentials with $D_e$ below this value are expected to be sensitive to the short-range part of the interaction potential. 
}
            \label{fig:de_dev}
        \end{figure}

To demonstrate the collision universality, and the breakdown thereof, we begin by comparing the results of rigorous quantum scattering calculations based on full interaction potentials with semi-classical calculations using Jeffreys-Born approximation (\ref{eq:JBJB}) parametrized only by the long-range part of the interaction potential (\ref{LR}). The calculations are performed for the collision systems and the conditions relevant for the vacuum metrology experiments
\cite{fagnanObservationQuantumDiffractive2009,
boothUniversalityQuantumDiffractive2019,barkerAccurateMeasurementLoss2023,klosElasticGlancingangleRate2023,shenCrosscalibrationAtomicPressure2023,shenRealizationUniversalQuantum2020,shenRefiningColdAtom2021,ehingerComparisonTwoMultiplexed2022,herpergerDeviationQuantumDiffraction2023,stewartMeasurementRbRbVan2022,boothRevisingUniversalityHypothesis2024}. Table~\ref{tab:q_sc_dev} lists the computed rate coefficients as well as the normalized difference:
    \begin{equation}
        \label{eq:def_uni}
        D = \frac{\kappa_\mathrm{Q} - \kappa_{\mathrm{SC}}}{\kappa_{\mathrm{SC}}}  \times 100\%,
    \end{equation}
    where Q denotes the quantum scattering calculations using the MLR fits (\ref{eq:mlr}) of the corresponding {\it ab initio} potentials and SC denotes the semi-classical calculations (\ref{eq:JBJB}). 
     It can be observed that $D$ for heavy background gas atoms (Ar, Kr, Xe) is much smaller than for light background gas atoms (He, Ne).  
      We note that the SC results based on Eq.~(\ref{eq:JBJB}) are fully parametrized by the long-range coefficients $C_n$ and completely ignore the short-range part of the interaction potential. 
The good agreement of the rigorous quantum scattering calculations and the SC results, therefore, shows that the rate coefficients for collisions with Ar, Kr, and Xe are predominantly determined by the long-range part of the interaction potential. 
Conversely, the lack of agreement between $\kappa_{\rm Q}$ and $\kappa_{\rm SC}$ indicates, though not proves, that the rate coefficients for collisions with He and Ne are sensitive to the short-range part of the interaction potential. 
The results of Table ~\ref{tab:q_sc_dev} thus indicate that the thermal collision rate coefficients at room temperature {are} universal for Ar, Kr, and Xe, but not for He or Ne.

        \begin{figure}[b]
             
                \includegraphics[width=0.45\textwidth]{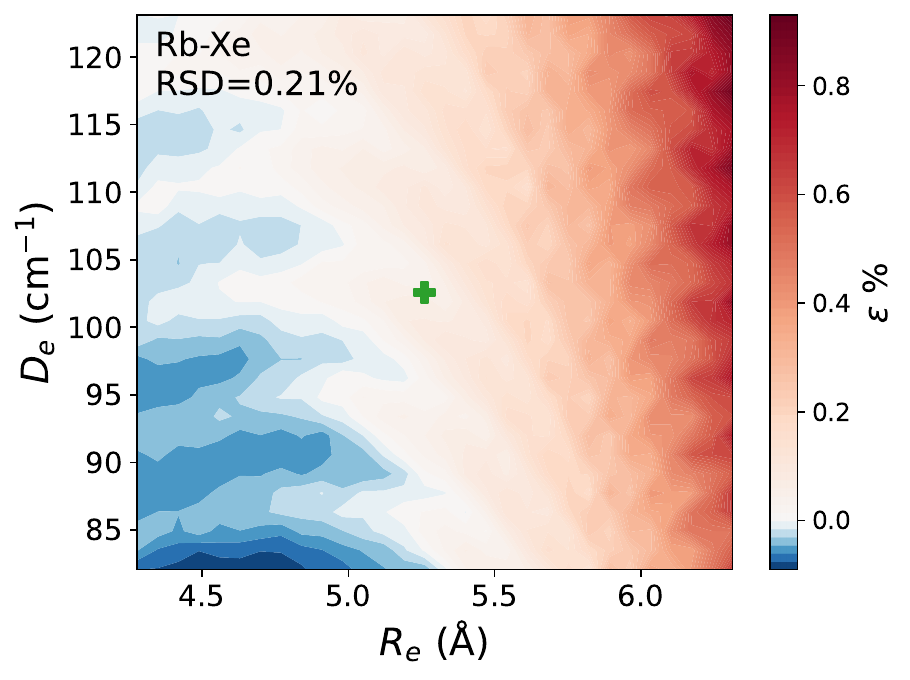} \\
                \includegraphics[width=0.45\textwidth]{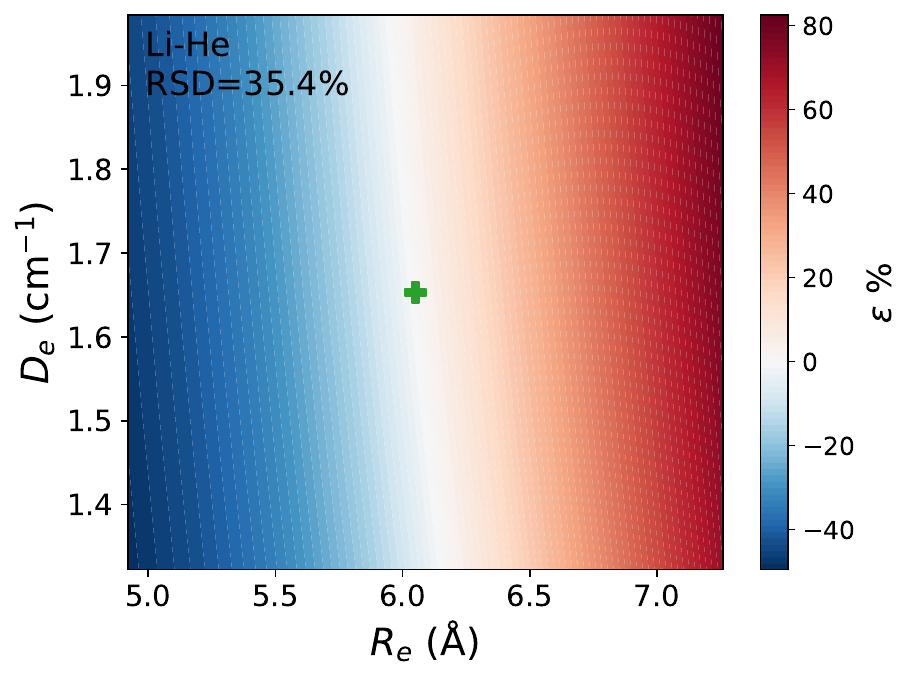}
            \caption{Relative errors (in \%) defined by Eq.~(\ref{eq:rel_err}) for Rb--Xe (upper panel) and Li--He (lower panel) collisions as functions of the well depth $D_e$ and the equilibrium distance $R_e$ of the corresponding interaction potentials. 
            The green plus symbol denotes the well depth $D_{e,0}$ and the equilibrium distance $R_{e,0}$ from the MLR fits (\ref{eq:mlr}) of the corresponding {\it ab initio} potentials for Rb -- Xe and Li -- He.  The values of $R_e$ and $D_e$ are sampled from the +/- 20\% intervals of $R_{e,0}$ and $D_{e,0}$ for each collision system.  RSD is the relative standard deviation of the rate coefficient distributions over these ranges of $R_e$ and $D_e$. }
            \label{fig:de_re_rate}
        \end{figure}

        \begin{figure}[b]
           
              \includegraphics[width=0.45\textwidth]{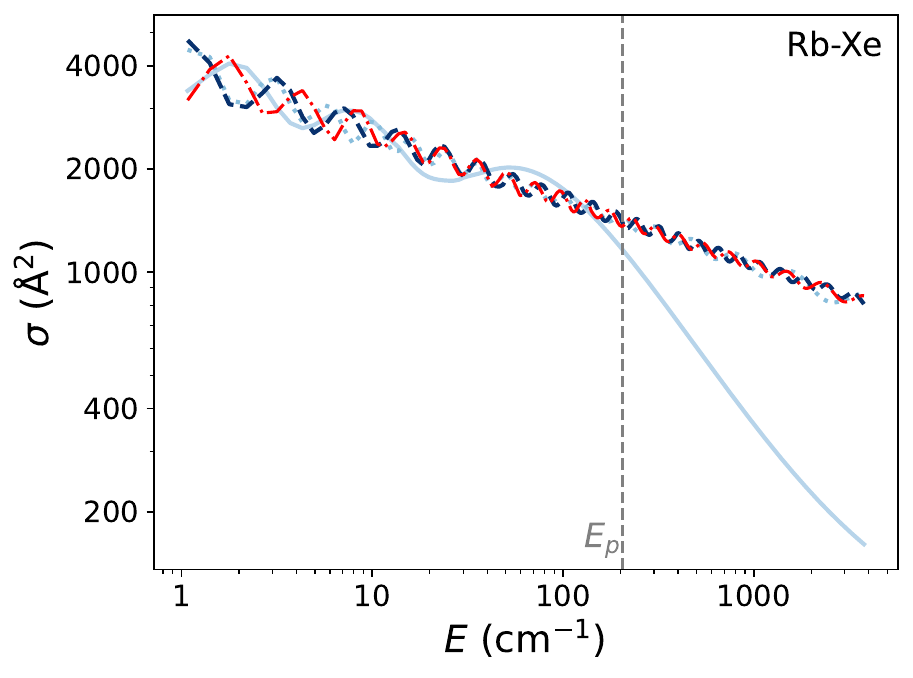}
              \includegraphics[width=0.45\textwidth]{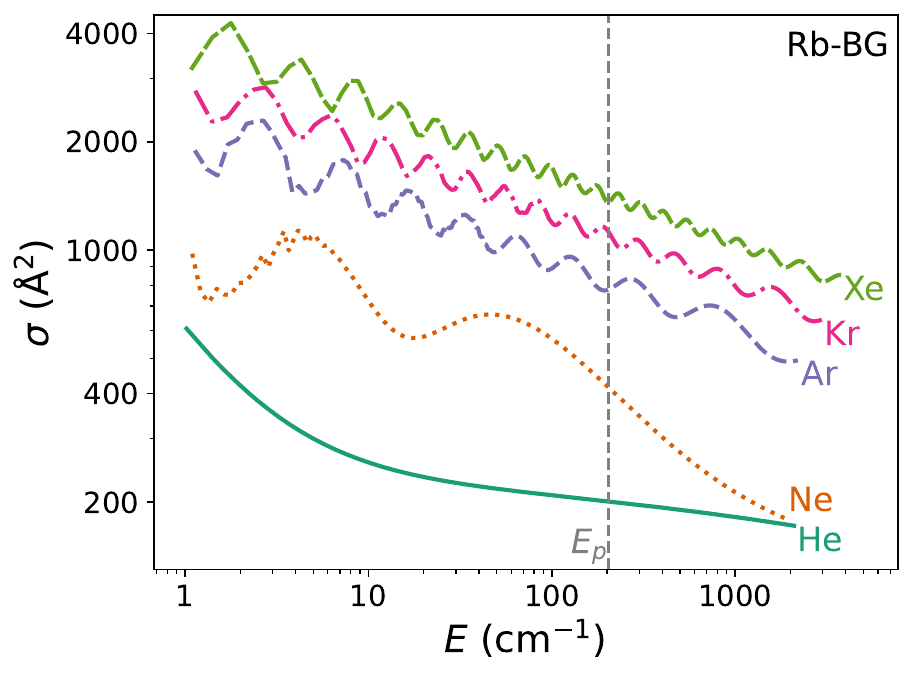}
          \caption{Upper panel: Energy dependence of the cross sections for Rb -- Xe collisions computed with four different interaction potentials parametrized by {$D_e = 1~\mathrm{cm}^{-1}$ (light blue, solid), 40~$\mathrm{cm}^{-1}$ (blue, dotted), 102.6~cm$^{-1}$ (red, dotdashed), 206~$\mathrm{cm}^{-1}$ (dark blue, dashed).} The red curve corresponds to the parameters of the best fit of the Rb -- Xe {\it ab initio} interaction potential.           
           The vertical grey dashed line represents $E_p = k_{\rm B} T = 204.44~\mathrm{cm}^{-1}$ at $T= 294.15$~K. Lower panel:  Energy dependencies of the cross sections for  Rb -- BG collisions: {BG = He (cyan-blue, solid), Ne (orange, dotted), Ar (purple, dashed), Kr (pink, dotdashed), Xe (green, densely dashed).} 
          }
          \label{fig:cole_crs}
      \end{figure}

        \begin{figure}[b]
              \includegraphics[width=0.45\textwidth]{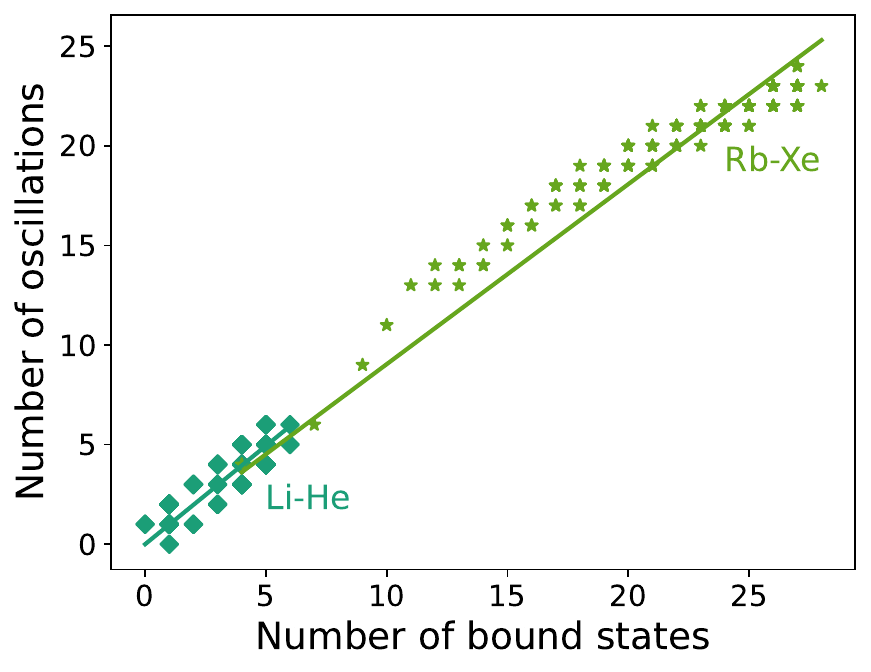}
              \includegraphics[width=0.45\textwidth]{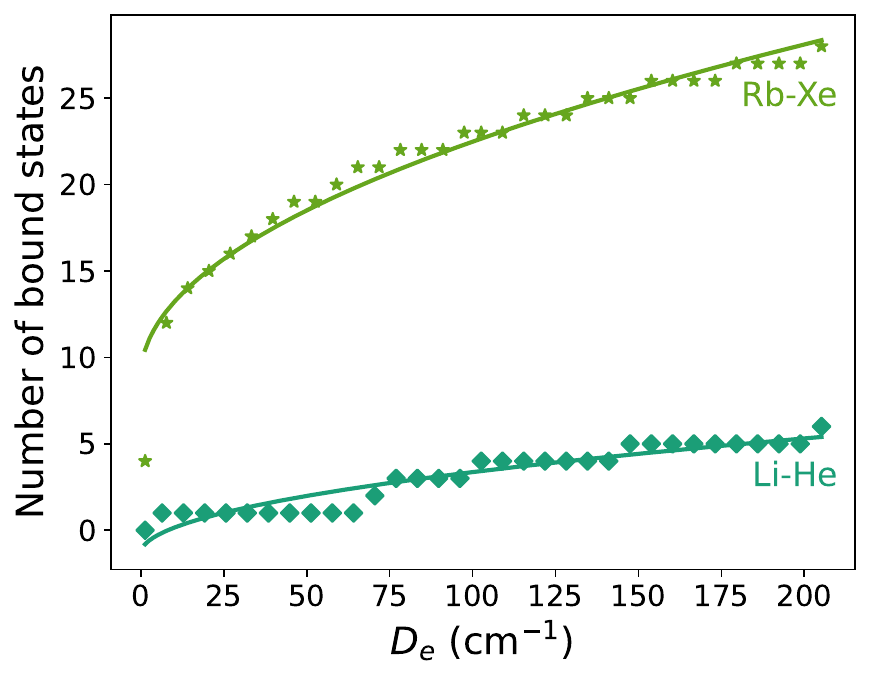}
          \caption{\label{fig:n_bound}   Number of oscillations in the energy dependence of the cross sections vs the number of bound states (upper panel) and $D_e$ (lower panel) for Rb -- Xe (green stars) and Li -- He (cyan-blue diamonds) interactions.  The solid lines are the linear fits (upper panel) and the fits to Eq.~(\ref{eq:n_osc}) in the lower panel. }
      \end{figure}

        \begin{figure}[b]
                \includegraphics[width=0.45\textwidth]{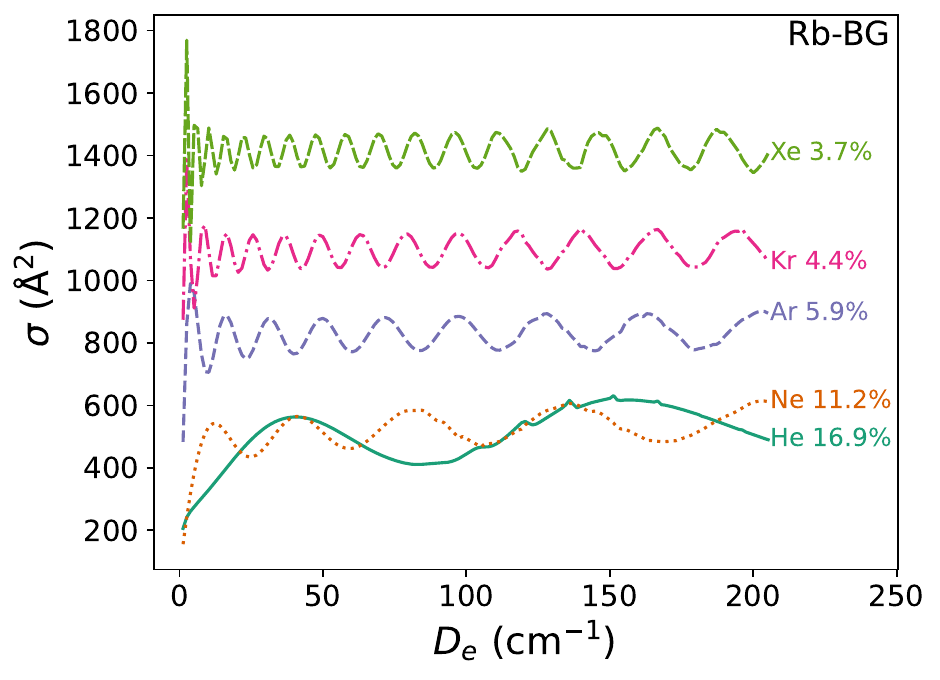}
                \includegraphics[width=0.45\textwidth]{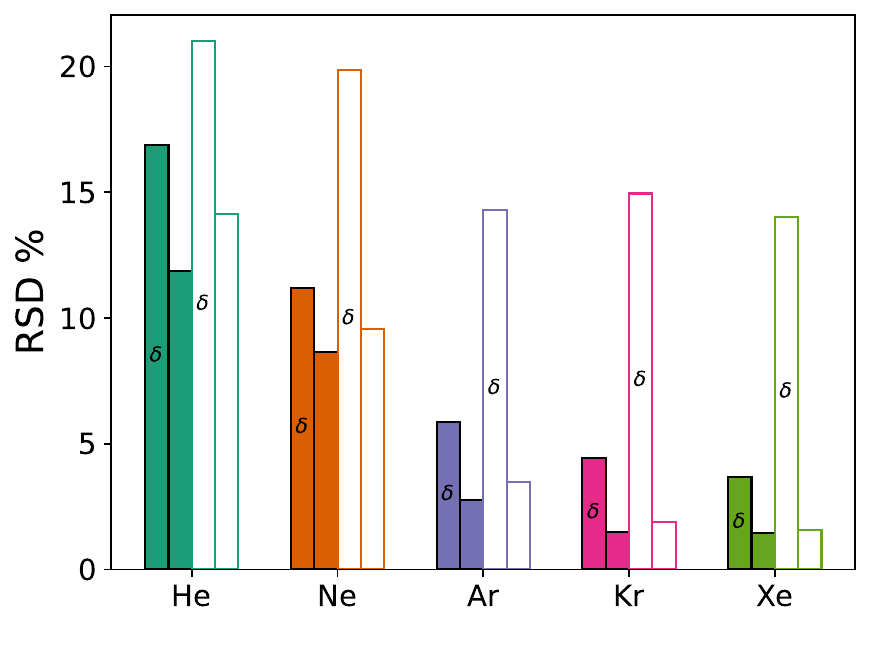}
            \caption{Upper panel: Cross sections for Rb-BG scattering as functions of $D_e$ at the energy $E_p = k_{\rm B} T= 204.44~\mathrm{cm}^{-1}$ corresponding to the maximum of the Maxwell-Boltzmann distribution at $T=294.15$ K for different BG gases: {He (cyan-blue, solid), Ne (orange, dotted), Ar (purple, dashed), Kr (pink, dotdashed), Xe (green, densely dashed).} The percentages are the relative standard deviation (RSD) of the distributions of $\sigma$ vs $D_e$. Lower panel: Relative standard deviation (RSD) of the distributions $f_\kappa(D_e)$ of the rate coefficients at $T = 294.15$ K for Rb-BG (full bars) and Li-BG (open bars) scattering. The bars labeled $\delta$ are computed with a delta function at the energy $E_p =k_{\rm B} T= 204.44~\mathrm{cm}^{-1}$ instead of the Maxwell-Boltzmann distribution. It should be noted that, as can be seen in the upper panel, the values of RSD are enhanced by the variation of the cross sections at small values of $D_e$.
            }
            \label{fig:de_crs_200_cm}
        \end{figure}

    \begin{figure*}
          \includegraphics[width=0.45\textwidth]{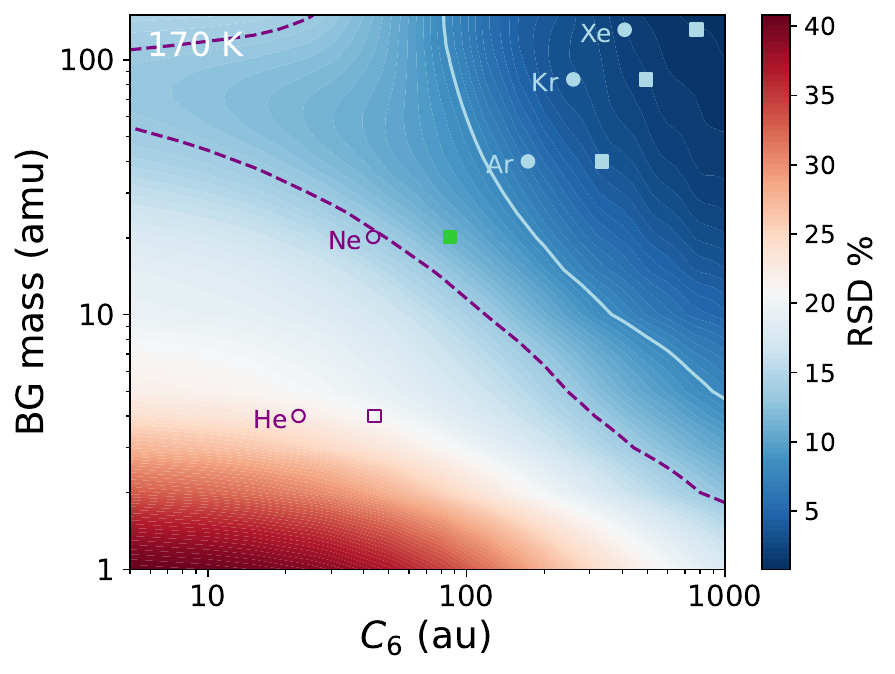}
          \includegraphics[width=0.45\textwidth]{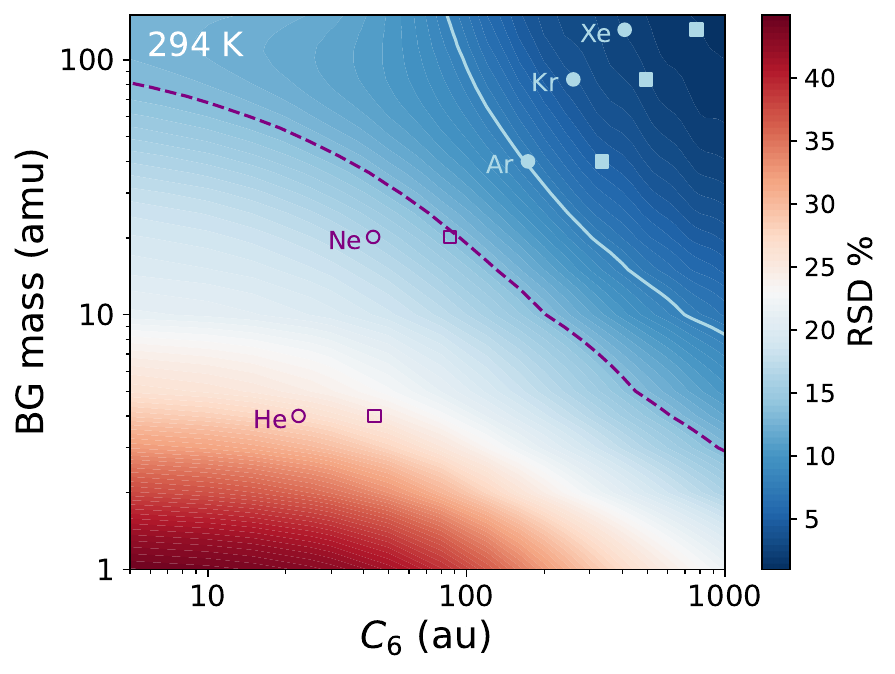}
          \includegraphics[width=0.45\textwidth]{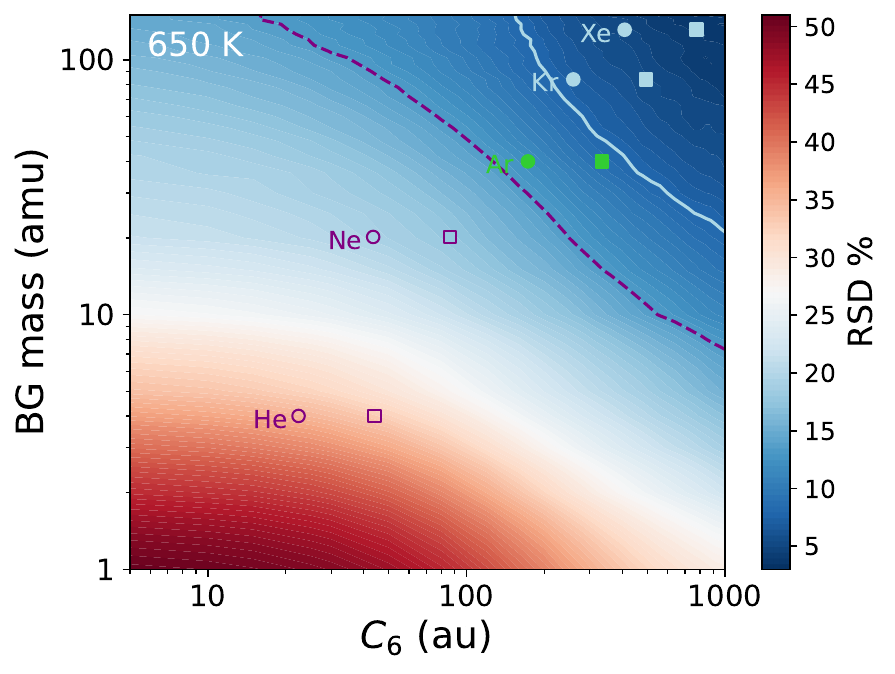}
          \includegraphics[width=0.45\textwidth]{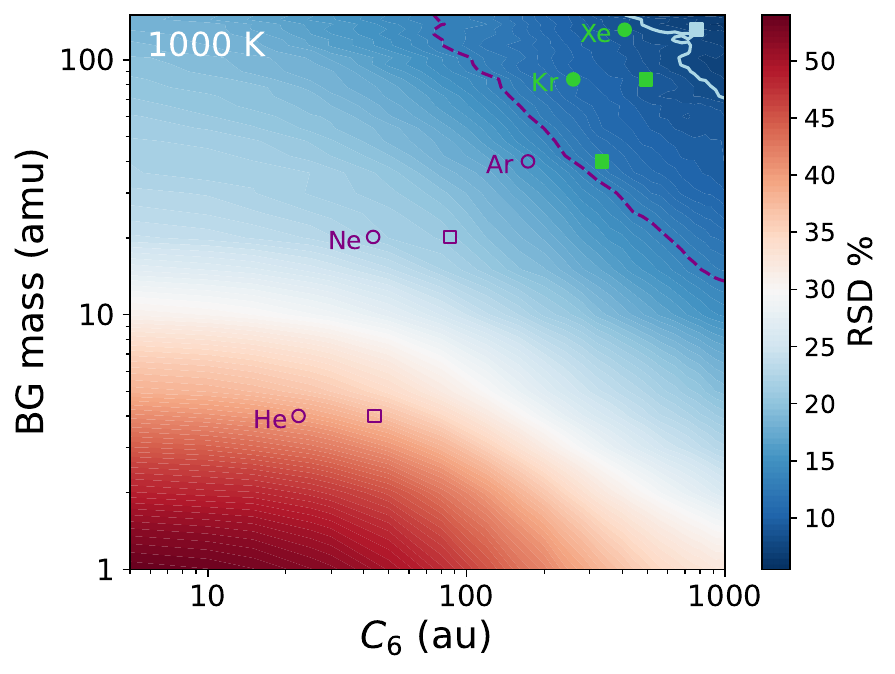}
      \caption{Relative standard deviation of the rate coefficient distributions as functions of the mass of the background gas atom and the value of $C_6$ for different temperatures indicated in the upper left corner of each panel. {All $C_6$ coefficients are in {atomic} units of ${E_{\rm h} \cdot a_0^6}$, where ${E_{\rm h}}$ is the Hartree energy and $a_0$ is the Bohr radius.} The {blue solid and purple dashed} lines represent  RSD = 8\% and 14\%, respectively. 
      The symbols show the parameters for the actual Rb-BG (squares) and Li-BG (circles) collision systems with BG = He, Ne, Ar, Kr, Xe. The full symbols show universal scattering, while the open symbols show non-universal collision systems. The green symbols show collision systems in the regime between the two boundaries.}
      \label{fig:bg_c6_var}
  \end{figure*}

  To verify this with rigorous quantum scattering calculations, we prepare a distribution of potential energy curves generated by varying 
  the well depth $D_e$ and the equilibrium distance $R_e$ of the interaction potentials (\ref{eq:mlr}) over the ranges $R_e \in [0.25,7.25]\,\text{\AA}$ and  $D_e \in [1,110]~\mathrm{cm}^{-1}$, yielding a total of 900 potentials for each collision system. These potentials are then used in quantum scattering calculations of the rate coefficients (\ref{eq:cal_rate}). Given these 900 collision rate coefficients, we train a two-dimensional quintic spline interpolator to generate the multivariate distribution $f_{\kappa}(D_e, R_e)$. We increase the number of grid points for interpolation until the convergence of $f_{\kappa}(D_e, R_e)$. The relative standard deviation (RSD) of the distribution for each collision system is presented in Table~\ref{tab:rate_distrib}. {To confirm that the set of 900 potentials accurately represents the distribution for each collision system, we have repeated the analysis using 3600 potentials, finding that the resulting RSDs change by less than 0.1\%.} 
  The results illustrate that RSD monotonically decreases with the mass of the collision partner for both Rb and Li sensor atoms. We also observe that RSD is significantly larger for He and Ne than for Ar, Kr, or Xe. The dependence of the room-temperature rate coefficients ratios ${\kappa (D_e)}/{\kappa_Q}$ on $D_e$ is further illustrated in Figure~\ref{fig:de_dev}.
   Combining the results of Tables~\ref{tab:q_sc_dev} and~\ref{tab:rate_distrib} and Fig.~\ref{fig:de_dev}, we classify Ar, Kr and Xe as universal and He and Ne as non-universal, at room temperature. This is consistent with the experimental observations 
 ~\cite{boothUniversalityQuantumDiffractive2019,shenRefiningColdAtom2021,shenCrosscalibrationAtomicPressure2023,boothRevisingUniversalityHypothesis2024,shenRealizationUniversalQuantum2020,herpergerDeviationQuantumDiffraction2023}.

  We note that the range of the interaction potentials considered to generate $f_{\kappa}(D_e, R_e)$ is much wider than the typical errors of quantum chemistry calculations~\cite{klosElasticGlancingangleRate2023,medvedevInitioInteratomicPotentials2018,blankNgPotentialEnergy2012}.  
  The range was deliberately exaggerated to examine the response of the thermal rate coefficients to variations in short-range interactions for both universal and non-universal collision systems. 
  To provide a more realistic illustration for the uncertainty of the collision rate coefficients due to possible errors of the {\it ab initio} calculations of the interaction potentials, we calculate the relative error
  {\begin{equation}
    \varepsilon  =  \frac{\kappa^\prime_{\mathrm{Q}} - \kappa_\mathrm{Q}}{\kappa_\mathrm{Q}} \times 100\%,
    \label{eq:rel_err}
\end{equation} 
where $\kappa_\mathrm{Q}$ are the collision rate coefficients listed in Table~\ref{tab:q_sc_dev} and $\kappa^\prime_\mathrm{Q}$}
are the rate coefficients from the quantum scattering calculations parametrized by the interaction potential with the well depth $D_e$ and the equilibrium distance $R_e$ sampled from ranges $R_e \in [0.8 \times R_{e,0},1.2 \times R_{e,0}]$ and $D_e \in [0.8 \times D_{e,0},1.2 \times D_{e,0}]$ centred at the fits of the corresponding {\it ab initio} potentials. Here,
 $R_{e,0}$ and $D_{e,0}$ are the well depth and the equilibrium distance characteristic of the individual collision system, obtained 
 from the MLR fits (\ref{eq:mlr}) of the corresponding {\it ab initio} potentials. 
 The values of $R_{e,0}$  and $D_{e,0}$, and consequently the ranges of $R_e$ and $D_e$, are thus different for the different collision systems. 
 The distribution of the relative errors as a function of $R_e$ and $D_e$ is shown in Figure~\ref{fig:de_re_rate}. 
 With the 20\% variation for the short-range part of the interaction potential, the thermal rate coefficients for Rb-Xe change by less than 1\%, while the same variation of the potentials leads to relative errors in the collision rate coefficients for Li-He reaching 81\%. 
Figure~\ref{fig:de_re_rate} also illustrates that the collision rate coefficients are more sensitive to variations in the equilibrium distance than in the well depth.

 To obtain more detailed information, we next examine the energy dependences of the cross sections for all collision systems involving Rb considered in this work. 
        We fix the equilibrium distance and well depth parameters to the values for the  MLR fit (\ref{eq:mlr}) of the corresponding {\it ab initio} potentials. In addition, for Rb -- Xe, we consider several interaction potentials with the well depth parameter in the range $D_e \in [1, 206]~\mathrm{cm}^{-1}$. 
        While $D_{e,0}$ for Rb -- Xe is 102.6~cm$^{-1}$, we aim to include an example of Rb -- Xe scattering with an artificial interaction potential resembling the interaction strength of Li and He.

        The upper panel of Figure~\ref{fig:cole_crs} shows the cross sections for Rb -- Xe collisions as determined by the interaction potentials with four values of the well depth parameter: $D_e = $ 1, 40, 102.6 ($D_{e,0}$) and 206~$\mathrm{cm}^{-1}$. 
        The energy dependences of the cross sections exhibit glory oscillations ~\cite{bernsteinSemiclassicalAnalysisExtrema1963a,bernsteinExtremaVelocityDependence1962,childMolecularCollisionTheory2014} in the energy range from 40 to 206~$\mathrm{cm}^{-1}$. 
        The corresponding collision rate coefficients are [2.02, 2.88, 2.89 and 2.90] $\times~10^{-15}~\mathrm{m^3/s}$. 
This illustrates that the integration over the Maxwell-Boltzmann distribution averages out the influence of the oscillations on the total rate coefficient. 
The outlier rate coefficient 2.02 $\times~10^{-15}~\mathrm{m^3/s}$ obtained with $D_e = 1$~cm$^{-1}$ indicates a transition to a non-universal scattering regime.

The lower panel of Figure~\ref{fig:cole_crs} compares the energy dependences of the cross sections for the different collision systems considered here. 
The results illustrate that the number of oscillations increases and the amplitude of the oscillations decreases with the reduced mass of the collision system~\cite{bernsteinExtremaVelocityDependence1962,bernsteinSemiclassicalAnalysisExtrema1963a,robertsGloryOscillationsIndex2002,childMolecularCollisionTheory2014,boothRevisingUniversalityHypothesis2024}. 
This increase in the frequency of the glory oscillations and the suppression of the amplitude reduce the effect of the differences in the cross section on the thermal rate coefficients, reducing the sensitivity of the collision rates to the short-range part of the interaction potentials~\cite{boothRevisingUniversalityHypothesis2024}. 

The energy range of the glory oscillations relative to the peak of the Maxwell-Boltzmann distribution determines the effect of thermal averaging. 
As shown in Ref.~\cite{boothRevisingUniversalityHypothesis2024}, the glory oscillations in $\sigma(E)$ extend to the energy $E_{\rm max} =  m (0.48 R_0 D_e/\hbar)^2/2$ where $R_0$ is the classical turning point {where} the interaction potential {energy equals zero}. The scattering cross section at $E > E_{\rm max}$ is predominantly determined by the short-range part of the interaction potential ~\cite{boothRevisingUniversalityHypothesis2024}. 
By equating $E_{\rm max}$ to $k_{\rm B} T$, this relation between $E_{\rm max}$ and $D_e$ can be used to estimate the values of $D_e$ that mark the transition between universal and non-universal scattering. The values of $D_e$ for $E_{\rm max} = k_{\rm B} T$ are illustrated by the open symbols in Figure~\ref{fig:de_dev}. 
The results in Figure~\ref{fig:de_dev} suggest that thermal collisions are expected to be sensitive to the short-range part of the interaction potential for $D_e$ below the value, where
the derivative of the thermal rate coefficients with respect to $D_e$ vanishes.

  The number of the glory oscillations in the energy dependence of the cross section is known to be approximately equal to the number of rotation-less bound states for the Lennard-Jones interaction potentials~\cite{childMolecularCollisionTheory2014,bernsteinExtremaVelocityDependence1962,bernsteinSemiclassicalAnalysisExtrema1963a}. 
 Since the present work employs more physical MLR functions (\ref{eq:mlr}) to model the interaction potentials, it is important to confirm the relation between the number of bound states and the oscillatory behavior of the energy dependence of the collision sections illustrated in Fig.~\ref{fig:cole_crs}.  
  We compute the bound states by diagonalizing the Hamiltonian matrix using the discrete variable representation approach of Colbert and Miller~\cite{millernNovelDiscreteVariable1992,kremsBreakingVanWaals2004} for each interaction potential used in the scattering calculations. 
  As shown in the upper panel of Figure~\ref{fig:n_bound}, the number of oscillations is a linear function of the number of rotation-less bound states. 
  The slopes of the solid lines in the upper panel of Figure~\ref{fig:n_bound} obtained from linear regression are 0.99 and 0.90, respectively. 
  
  To obtain the dependence of the oscillatory behavior of the cross sections on $D_e$, 
we use this linear regression along with the semi-classical approximation based on the generalized Bernstein function~\cite{bernsteinSemiclassicalAnalysisExtrema1963a}
        \begin{equation}
            N_{\rm osc} = N_{\rm bound} = \alpha \sqrt{ \frac{2 \mu R_0^2 D_e }{\hbar^2}} + \beta,
            \label{eq:n_osc}
        \end{equation} where $N_{\rm osc}$ is the number of oscillations in the energy dependence of the cross section, $N_{\rm bound}$ is the number of rotation-less bound states in the corresponding interaction potential, $\mu$ is the reduced mass, and $R_0$ is the interatomic separation where the interaction potential vanishes. 
        The dimensionless parameters $\alpha$ and $\beta$ are obtained by non-linear least squares fitting. The resulting fits (\ref{eq:n_osc}) are shown in the lower panel of Figure~\ref{fig:n_bound}.         
                The outlier for the Rb -- Xe fit corresponds to the outlier curve in the upper panel of Figure~\ref{fig:cole_crs}, as well as the outlier rate 2.02 $\times~10^{-15}~\mathrm{m^3/s}$, which again indicates the universality break-down at the well depth parameter  $D_e = 1~\mathrm{cm}^{-1}$. 
                This illustrates that interaction potentials supporting a large number of bound states and a large reduced mass favor the scattering regime that reduces the sensitivity of thermal collision rate coefficients to the short-range part of the interaction potential. 
                
        To illustrate the change of the amplitude of the glory oscillations with the collision partner in more detail, we calculate the cross sections at the energy $E_p =k_{\rm B} T=204.44~\mathrm{cm}^{-1}$ as functions of $D_e$. The results are shown in the upper panel of Figure~\ref{fig:de_crs_200_cm}. The cross sections as functions of $D_e$ exhibit oscillations, whose frequency decreases with increasing $D_e$. This 
        corresponds to the diminishing rate of the increase of the number of rotation-less bound states as the well depth grows. 
The effect of the amplitude of the glory oscillations on thermal rate coefficients can be demonstrated by the relative standard deviation (RSD) of the distribution of the cross sections vs $D_e$. The values of RSD are shown as percentages in the upper panel of Figure~\ref{fig:de_crs_200_cm}. 
                These values of RSD quantify the uncertainty of the oscillating cross sections relative to their means, which is equivalent to the uncertainty of the rate coefficients when the {energy distribution of $E$} is a delta function at $E_p$. 
                Thus, the RSD provides an estimate of the relative residual effect of the oscillation amplitude on the rate coefficients. 
                
                Without the averaging effect of the Maxwell-Boltzmann distribution, the values of RSD represent the upper bound of the uncertainty in the thermal rate calculation due to the uncertainty of the well depth of the interaction potential. 
                To illustrate how Maxwell-Boltzmann averaging reduces the sensitivity to the glory oscillations, we compare the RSD values computed with and without the Maxwell-Boltzmann averaging for both Rb -- BG and Li -- BG collisions. The results are shown in the lower panel of Figure~\ref{fig:de_crs_200_cm}. %
With a heavier background gas, the amplitudes of the glory oscillations at the energy $E_p = k_{\rm B} T$ become lower, leading to a lower upper bound of the uncertainty. For each background gas, the difference between the bars with and without the $\delta$ label shows the reduction of the uncertainty upper limit due to integration over the Maxwell-Boltzmann distribution. 
        
        The universality can thus be viewed as a consequence of the thermal averaging that diminishes the effect of the glory oscillations
        appearing in the energy dependence of the cross section due to the short-range part of the interaction potential. There are three factors that contribute to the onset of universality: the energy range of the glory oscillations relative to the maximum and width of the Maxwell-Boltzmann distribution, the number  and the amplitude of the oscillations. 
                For a given collision system within the universal scattering regime, the amplitude of glory oscillations determines the upper bound of the sensitivity to the short-range part of the interaction potential, while the number of glory oscillations determines the reduction of the uncertainty due to integration over the Maxwell-Boltzmann distribution. 
        The number of the oscillations is determined by the number of bound states in the corresponding interaction potential. The Maxwell-Boltzmann distribution is determined by temperature and the mass of the background gas atoms. 
        Whether a thermal rate coefficient is sensitive to the short-range part of the interaction potential is thus determined by the following factors: the binding energy of the interaction potential, the mass of the background gas atoms, temperature and the polarizability of the interacting atoms as measured by the long-range $C_6$ coefficient. We note that the binding energy $D_e$ and the long-range interaction coefficients $C_6$ are not independent as atomic interactions with large $C_6$ generally have large $D_e$. In the following section, we examine the boundaries of the universality in the $[C_6, m, T]$ parameter space using distributions of thermal rate coefficients with $D_e$ and $R_e$ treated as random variables.

    \subsection{Boundaries of collision rate universality} 
            \label{sec:limit_uni}

    To provide guidance for experimental measurements using universality, we explore the boundaries of universal scattering as functions of the dipole-dipole interaction coefficient $C_6$, masses of background gas atoms, and temperature. 
    We consider Li as the trapped atom and a range of artificial collision systems with the following parameters: 
    $m_{bg} \in [1, 150]~\text{amu}$ and $C_6 \in [10, 1000]~{E_{\rm h}\cdot a_0^6}$. 
We then perform quantum scattering calculations for each artificial collision system to obtain the RSD of the distribution $f_\kappa(D_e, R_e)$ using the approach described in Section~\ref{sec:ill_uni}. 
To illustrate how the regime of universal scattering changes with temperature, we plot RSD of the distributions $f_\kappa(D_e, R_e)$ as functions of $C_6$ and the mass of background gas atom at $T = 170, 294, 650$, and $1000$~K in Figure~\ref{fig:bg_c6_var}.
Guided by the results in Table~\ref{tab:rate_distrib}, we consider RSD $<$ 8\% to indicate universal scattering, and $>$ 14\% to indicate non-universal scattering. These boundaries are shown by the lines in Figure~\ref{fig:bg_c6_var}. 
To produce definitive conclusions, we treat the results with  $8\% <$ RSD $< 14\%$ as borderline. 
We use the same values of RSD at different temperatures to examine the shift of these boundaries with $T$.

    It can be observed that the regime of universal scattering 
    shrinks as temperature increases. This reflects the effect of the broadening of the Maxwell-Boltzmann distribution, allowing the distribution, {along with its center,} to reach beyond the energy range of the glory oscillations. At the same time, Rb-Xe scattering remains universal up to $T=1000$~K. In general, it can be seen that a larger value of $D_e$ and a larger mass of the background gas atom leads to weaker sensitivity of the rate coefficients to variations in the short-range part of the interaction potential. {The universal regime for collisions at room temperature occurs when contributions to the energy-averaged rate coefficient are negligible for $E$ values exceeding $E_{\text{max}}=m(0.48 R_0 D_e/\hbar)^2/2$~\cite{boothRevisingUniversalityHypothesis2024}, where cross sections are influenced primarily by the short-range part of interaction potential. This suggests that light background species at higher temperatures do not exhibit universality, as the Maxwell-Boltzmann distribution has substantial weight at energies above $E_{\text{max}}$.}

It is important to quantify the residual sensitivity of rate coefficients to the short-range part of the interaction potentials for thermal collisions within the regime of universal scattering.  
Table~\ref{tab:rate_distrib} shows RSD$^\ast$ of the thermal rate coefficient distributions induced by a distribution of interaction potentials covering a $\pm 20\%$ range of $D_e$ and $R_e$ values for each collision system. This variation of interaction potentials significantly exceeds the expected uncertainty of the {\it ab initio} calculations and fits of the interaction potentials~\cite{klosElasticGlancingangleRate2023}. The values of RSD$^\ast$ reported in Table~\ref{tab:rate_distrib} for Ar, Kr and Xe can thus be viewed as the upper limits of the uncertainties of the rate coefficients at room temperature for the fixed values of the long-range interaction parameters $C_n$. If the ranges of $D_e$ and $R_e$ are reduced to $\pm~10\%$, the values of RSD$^\ast$ reduce to 0.1\% (for Rb-Xe and Li-Xe), 0.2\% (for Rb-Kr and Li-Kr), 0.5\% (for Rb-Ar) and 0.7\% (for Li - Ar). {We observe that RSD$^\ast$ for Rb-He in Table~\ref{tab:rate_distrib} exceeds its RSD. This is because Rb-He has the largest equilibrium distance among the systems considered in this work, and, as noted earlier, the rate coefficients are more sensitive to variations in the equilibrium distance than in the well depth. Moreover, the rate coefficient for Rb-He exhibits more rapid variation near its well depth compared to other collision systems, as shown in Fig.~\ref{fig:de_dev}.}

{We note that the electronic structure calculations for light atoms (e.g. He, Ne) are more tractable than those for heavy atoms (e.g. Ar, Kr, Xe). Consequently, the uncertainties in interaction potentials for light atoms are smaller than those for heavy atoms~\cite{klosElasticGlancingangleRate2023,medvedevInitioInteratomicPotentials2018}. This implies that for non-universal systems, the comparatively low uncertainties in interaction potentials provide reliable rate coefficient predictions despite the lack of universality. The numerical results in supplemental material~\cite{supplemental} illustrate the uncertainties of collision rate coefficients with typical uncertainties in interaction potentials from quantum chemistry calculations.}

   \subsection{Experiment - theory discrepancy}
     
     \begin{figure}[b]   
          \includegraphics[width=0.45\textwidth]{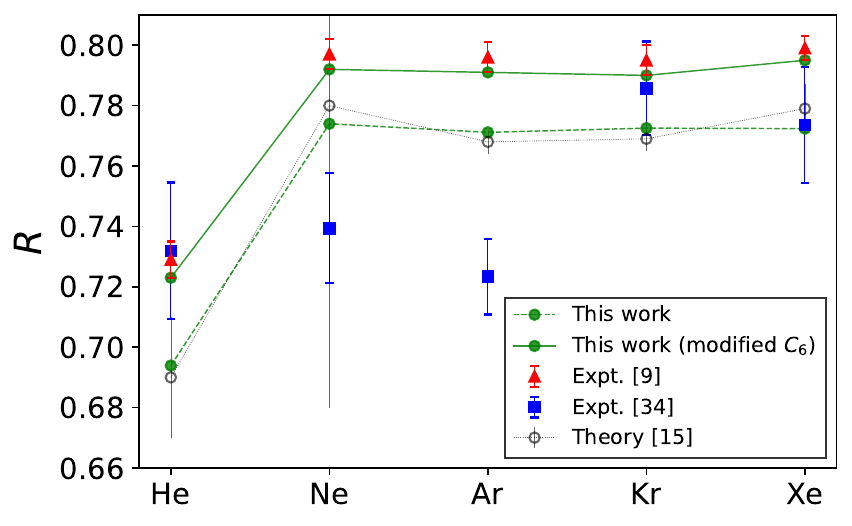}
      \caption{Ratios (\ref{eq:ratio}) of rate coefficients from different sources: squares from Ref.~\cite{barkerAccurateMeasurementLoss2023} and open circles from Ref.~\cite{klosElasticGlancingangleRate2023}; triangles from Ref.~\cite{frielingCrosscalibrationQuantumAtomic2024} and full circles with lines from the present calculations with the unmodified ({dashed} line)
     and modified (solid line) values of the $C_6$ coefficients.  
      }
      \label{fig:rb_li_rate_ratio}
      \end{figure}

As shown in Figure~\ref{fig:rb_li_rate_ratio},  
    the most recent experimental measurements~\cite{frielingCrosscalibrationQuantumAtomic2024} of the following ratio of the rate coefficients (at 298 K)
    \begin{equation}
        R = \frac{\kappa_{\mathrm{Li-BG}}}{\kappa_{\mathrm{Rb-BG}}}
        \label{eq:ratio}
    \end{equation}
      differ systematically (by up to 3\%) from the theoretical predictions in Ref.~\cite{klosElasticGlancingangleRate2023}, even for the heavy BG atoms Ar, Kr, and Xe expected to exhibit universal scattering. The difference is even larger (up to 8\%), compared to the ratios from the experimental measurements in Ref.\cite{barkerAccurateMeasurementLoss2023}. 

     The experiments in Ref.~\cite{barkerAccurateMeasurementLoss2023} and Ref.~\cite{frielingCrosscalibrationQuantumAtomic2024}  use different approaches. Thus, the experiments in Ref.~\cite{barkerAccurateMeasurementLoss2023} produce measurements of the rate coefficients for Li and Rb using two different apparatuses, whereas the experiment in Ref.~\cite{frielingCrosscalibrationQuantumAtomic2024}  exposes two traps, one with Rb and one with Li, to the same ambient gas. While we do not address the discrepancy between the experimental measurements in this work, we examine the implications of the discrepancy between the measurements in Ref.~\cite{frielingCrosscalibrationQuantumAtomic2024} and the results of the theoretical calculations. { 
        The interpretation of results from experimental measurements in Ref.~\cite{barkerAccurateMeasurementLoss2023} did not consider the effects of glancing angle collisions. It was shown previously~\cite{eckel2024effect} that glancing angle collisions, which do not immediately result in particle loss, alter the energy distribution of atoms in a trap. Therefore, an accurate interpretation of results from experimental measurements of the total collision rate must account for the evolution of the ensemble energy~\cite{Avinash2024}. These effects were considered in Refs.~\cite{frielingCrosscalibrationQuantumAtomic2024,eckel2024effect}. We use the rate coefficients from Refs.~\cite{frielingCrosscalibrationQuantumAtomic2024,eckel2024effect} in the present analysis.}

        We repeated the calculations of Ref.~\cite{klosElasticGlancingangleRate2023} using the interaction potentials from Ref.~\cite{klosElasticGlancingangleRate2023} and obtained the rate coefficients in agreement with the theoretical results of  Ref.~\cite{klosElasticGlancingangleRate2023}, as shown 
              in Figure~\ref{fig:rb_li_rate_ratio}.    This confirms the consistency of both the current calculations and the calculations in Ref.~\cite{klosElasticGlancingangleRate2023} for the specific interaction potentials employed in Ref.~\cite{klosElasticGlancingangleRate2023}. These interaction potentials used the values of $C_6$ from Ref.~\cite{dereviankoElectricDipolePolarizabilities2010} summarized in Table~\ref{tab:c6_c6_exact}. We now show that a $\sim 4\%$ modification of the values of these $ C_6$ coefficients can bring the results of the theoretical calculations in alignment with the lower limit of the experimental measurements in Ref.~\cite{frielingCrosscalibrationQuantumAtomic2024}. 
       
\begin{figure*}
              \includegraphics[width=0.45\textwidth]{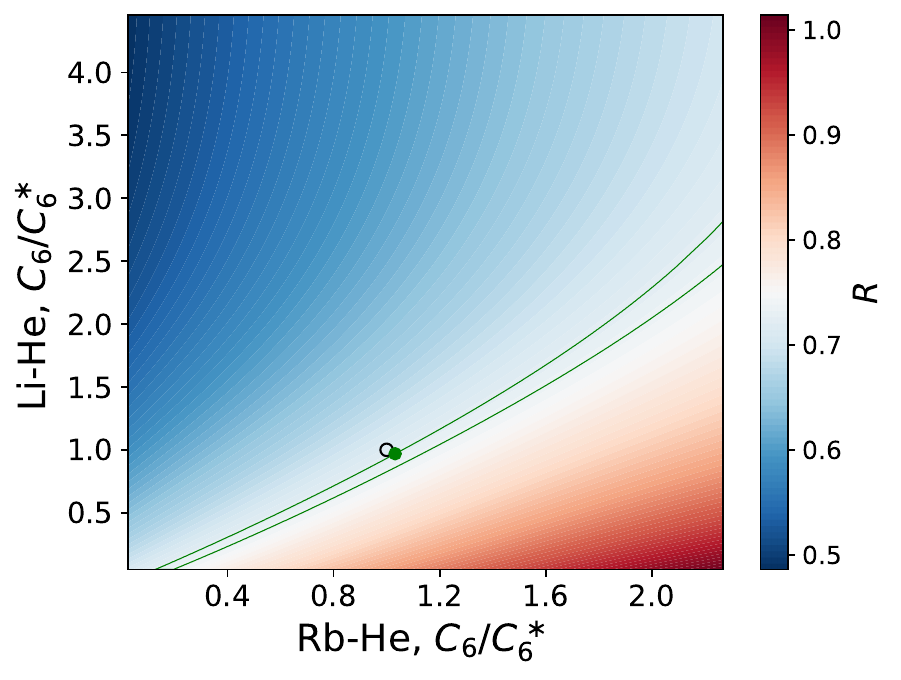}
              \includegraphics[width=0.45\textwidth]{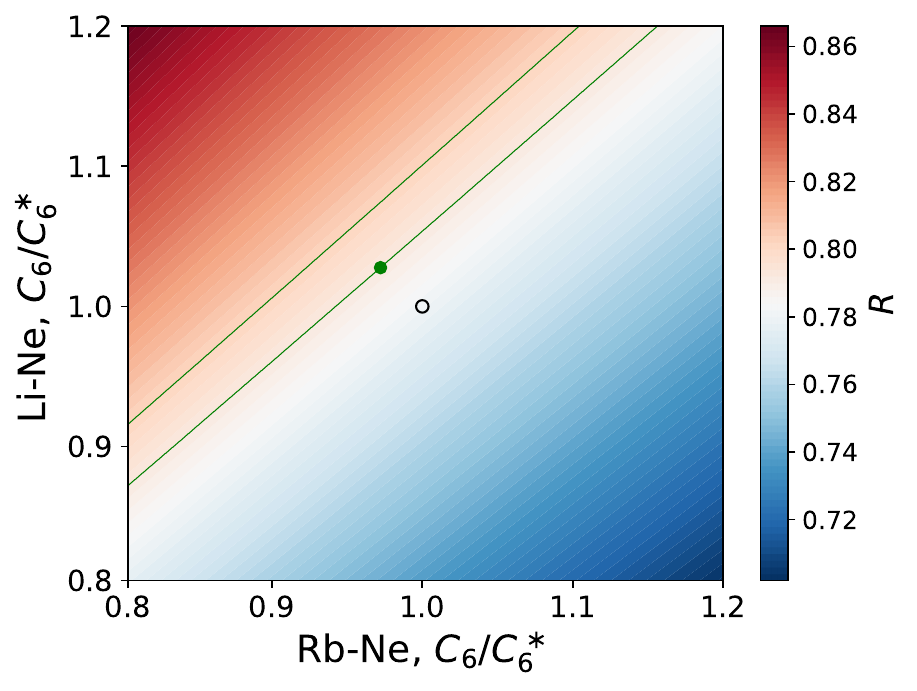}
              \includegraphics[width=0.45\textwidth]{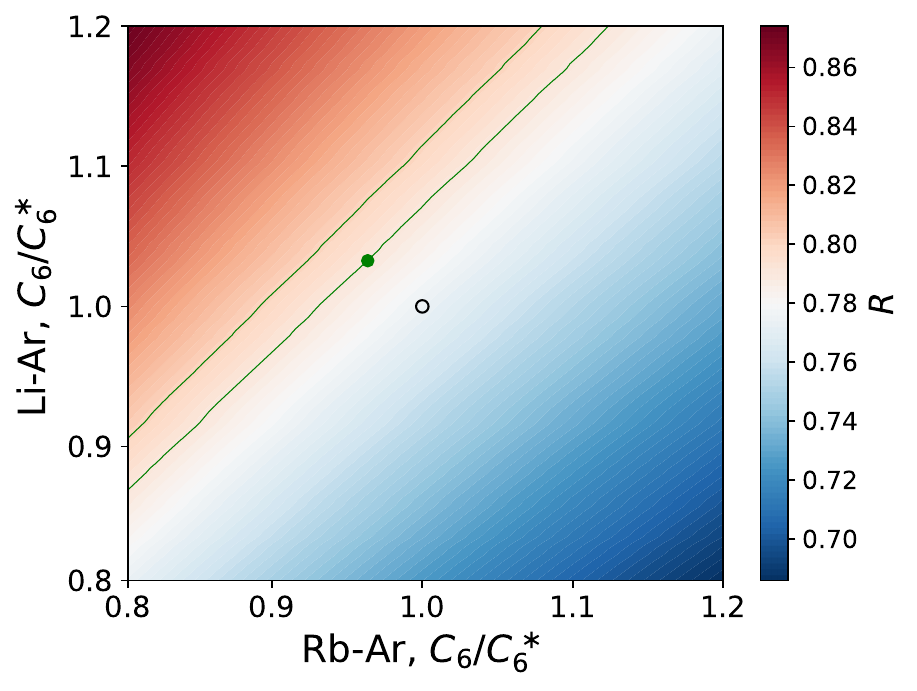}
              \includegraphics[width=0.45\textwidth]{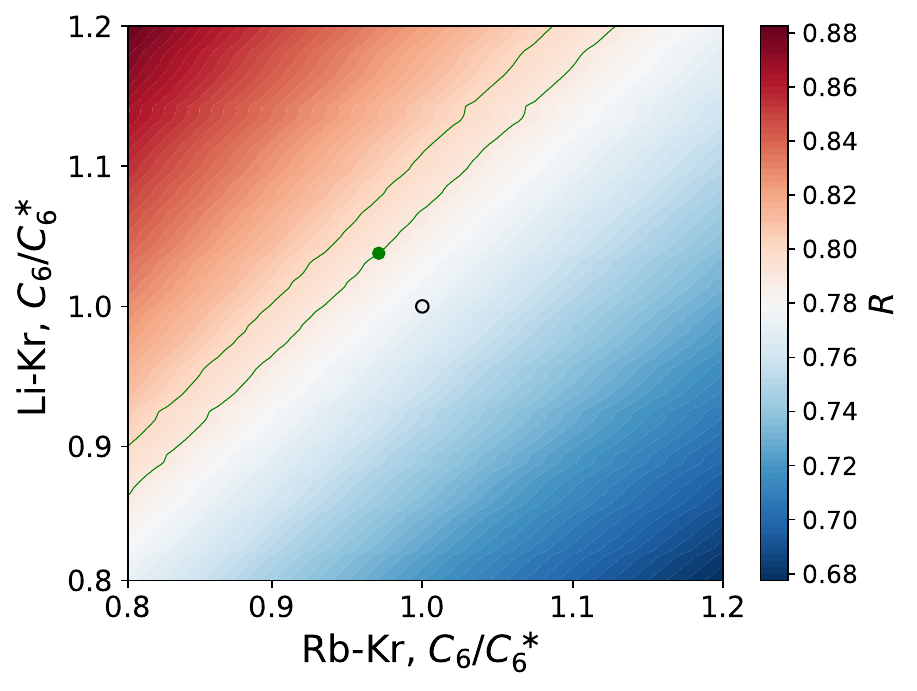}
              \includegraphics[width=0.45\textwidth]{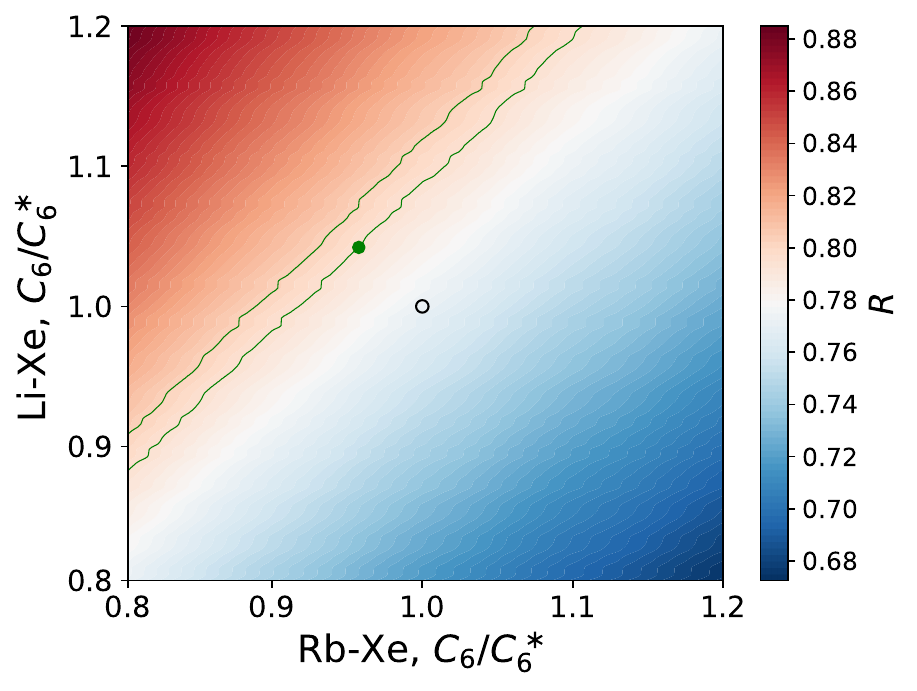}
  
          \caption{Ratios (\ref{eq:ratio}) of rate coefficients for  Rb-BG and Li-BG scattering as functions of ${C_6}/{C^\ast_{6}}$, where ${C^\ast_{6}}$ is the dipole-dipole interaction coefficients from Ref~\cite{klosElasticGlancingangleRate2023}. 
         The hollow circles indicate the combinations of ${C^\ast_{6}}$ for the corresponding two systems. 
        The solid lines show the region where the ratios (\ref{eq:ratio})  fall within the error bars of the measurements in Ref.~\cite{frielingCrosscalibrationQuantumAtomic2024}. 
        The green circles show the point in the region between the dashed lines that has the shortest Euclidean distance to the values of ${C^\ast_{6}}$ (shown by the circles).}
          \label{fig:c6_c6_ratio_exact}
      \end{figure*}

       Section~\ref{sec:ill_uni} illustrates that the rate coefficients for collisions of Li and Rb with Ar, Kr, and Xe are not sensitive to variations of the short-range part of the interaction potential. 
       This leaves the uncertainty in the values of the long-range interaction parameters as the only possible source of the uncertainty of the quantum scattering calculations. 
       To examine the sensitivity of the rate coefficient to the long-range dispersion interaction, we fix the short-range part of each interaction potential to the best fit of {\it ab initio} results and 
       vary the $C_6$ coefficients for both Li-BG and Rb-BG systems (BG = He, Ne, Ar, Kr, Xe) in a range centered around the values from Ref.~\cite{klosElasticGlancingangleRate2023}, denoted in Fig.~\ref{fig:c6_c6_ratio_exact} by $C_6^\ast$. 
 For each combination of the $C_6$ coefficients, we calculate the ratio (\ref{eq:ratio}) of the rate coefficients. The results are presented in Fig.~\ref{fig:c6_c6_ratio_exact} for all collision systems. The solid lines in Fig.~\ref{fig:c6_c6_ratio_exact} show the range of the $C_6$ values that produce the ratio (\ref{eq:ratio})  within the error bars of the experimental measurements~\cite{frielingCrosscalibrationQuantumAtomic2024}. We note that the solid lines follow the relation 
 \begin{eqnarray}
    C_6^{\text{Li-BG}}= R^{\frac{5}{2}}C_6^{\text{Rb-BG}}
 \end{eqnarray} 
 for BG = Ne, Ar, Kr, Xe~\cite{frielingCrosscalibrationQuantumAtomic2024}. For each background gas atom, a green circle in Fig.~\ref{fig:c6_c6_ratio_exact} shows the point in the region between the solid lines that requires the smallest corrections to the reported $C_6$ values (shown by the hollow circles) used in Ref.~\cite{klosElasticGlancingangleRate2023}. {We note that the scale of the panel for He is distinct from that of other gases to better illustrate the nonlinearity of the solid lines in the panel for He.}
The values of the $C_6$ coefficients represented in Fig.~\ref{fig:c6_c6_ratio_exact} by the green circles are listed in Table~\ref{tab:c6_c6_exact} and used for the calculations producing the upper green line in Fig.~\ref{fig:rb_li_rate_ratio}, in agreement with the experimental measurements~\cite{frielingCrosscalibrationQuantumAtomic2024}. 
 
The results shown in Fig.~\ref{fig:c6_c6_ratio_exact} and Table~\ref{tab:c6_c6_exact} thus illustrate that a change of $C_6$ coefficients by $< 4.2\%$ is sufficient to align the quantum scattering calculations with either of the two experimental measurements. This indicates that the experiments in {Refs.~\cite{barkerAccurateMeasurementLoss2023,eckel2024effect}} and~\cite{frielingCrosscalibrationQuantumAtomic2024} can be used to constrain the values of the $C_6$ coefficients. We note that the agreement of the theoretical calculations with the modified $C_6$ coefficients and the measurements in Ref.~\cite{frielingCrosscalibrationQuantumAtomic2024}  is better than the agreement between the calculations in Ref.~\cite{klosElasticGlancingangleRate2023} and measurements in {Refs.~\cite{barkerAccurateMeasurementLoss2023,eckel2024effect}}, which shows a significant separation between the theoretical and experimental results for Ne and Ar. 
        
    \begin{table}[b]    
      \caption{\label{tab:c6_c6_exact}
  Values of $C_6$ coefficients, originally from Ref.~\cite{dereviankoElectricDipolePolarizabilities2010}, employed for the calculations in Ref.~\cite{klosElasticGlancingangleRate2023} and in this work. All the $C_6$ coefficients are in {atomic} units of ${E_{\rm h} \cdot a_0^6}$, where ${E_{\rm h}}$ is the Hartree energy and $a_0$ is the Bohr radius. The $C_6$ coefficients from this work correspond to the full  circles in Fig.~\ref{fig:c6_c6_ratio_exact}.}
      \begin{ruledtabular}
      \begin{tabular}{ccccc}
      Sensor & BG & Refs.~\cite{klosElasticGlancingangleRate2023, dereviankoElectricDipolePolarizabilities2010} & This work & Difference \% \\ 
      \hline
      Rb & He & 44.07 & 45.48 & 3.2 \\ 
      Li & He & 22.44 & 21.74 & -3.1 \\ 
      Rb & Ne & 86.50 & 84.10 & -2.8 \\ 
      Li & Ne & 43.60 & 44.81 & 2.8 \\ 
      Rb & Ar & 334.0 & 321.89 & -3.6 \\ 
      Li & Ar & 173.0 & 178.64 & 3.3 \\ 
      Rb & Kr & 494.0 & 479.70 & -2.9 \\ 
      Li & Kr & 259.0 & 268.83 & 3.8 \\ 
      Rb & Xe & 776.0 & 743.25 & -4.2 \\ 
      Li & Xe & 409.0 & 426.22 & 4.2 \\ 
      \end{tabular}
      \end{ruledtabular}
      \end{table}

\section{Conclusion}

We have presented a comprehensive analysis of the response of thermally averaged rate coefficients for atom - atom collisions to variations in the short-range part of the interaction potentials as well as the dispersion interaction parametrized by $C_6$.  
We use rigorous quantum scattering theory and interaction potentials based on high-level electronic structure calculations to examine in detail the mechanism of collision universality that manifests itself in the invariance of thermal collision rates to changes in the short-range part of the interaction potentials. 
It is illustrated that collision universality is a result of the averaging over the Maxwell-Boltzmann distribution that reduces the effect of short-range interactions leading to glory oscillations in the energy dependence of the cross sections. Our scattering calculations systematically illustrate that 
the number of the oscillations increases with the number of bound states supported by the corresponding interaction potential and the amplitude of the oscillations decreases with the reduced mass of the collision complex. It is thus illustrated that heavy polarizable species, characterized by interaction potentials supporting a large number of bound states,  are more likely to exhibit universal scattering than light, few-electron atoms. 

In order to guide future experiments aiming to exploit the universality of thermal rates, we examine the variance of collision rate coefficients over the distribution of interaction potentials with widely varying short-range behavior, at different temperatures. Our calculations produce diagrams that identify regions of universality, as well as the lack thereof, in the [atom mass, $C_6$, $T$] parameter space. Given long-range interaction parameters, any collision system can be identified with a region in these diagrams. 

Our analysis of the sensitivity of thermal collision rate coefficients to the long-range dispersion interaction parameter $C_6$ indicates that measurements of trap loss due to ambient pressure in an ultra-high-vacuum regime can be used as sensitive probes of long-range interactions. We have shown that a 3\% difference in the measurement of rate coefficient ratios discriminates $C_6$ values at the 4\% level. 
We note that this change of the $C_6$ coefficients exceeds the uncertainties of the $C_6$ coefficients estimated in Ref.~\cite{dereviankoElectricDipolePolarizabilities2010}
by using experimental error bars for principle transitions and by comparison of the atom - wall interaction constants obtained by integrating the corresponding dynamic polarizabilities. 
Our present results call for reexamination of these uncertainties of the $C_6$ coefficients in Ref.~\cite{dereviankoElectricDipolePolarizabilities2010} or for a careful examination of the effect of higher-order $C_8$ and $C_{10}$ coefficients on thermal collision rates. 
Our results underscore the importance of high-precision calculations of long-range interaction parameters for vacuum metrology with trapped atoms. For example, reducing the error bars on the calculated $C_6$ coefficients can be used to validate the experimental measurements, especially for heavy ambient gas atoms.

\section*{Acknowldgments}

We acknowledge the financial support from the Natural Sciences and Engineering Research Council of Canada (NSERC grants RGPIN-2019-04200, RGPAS-2019-00055) and the Canadian Foundation for Innovation (CFI project 35724). This work was done at the Center for Research on Ultra-Cold Systems (CRUCS) and was supported in part, through computational resources and services provided by Advanced Research Computing at the University of British Columbia.  X.G., K.W.M. and R.V.K. acknowledge support from the Deutsche Forschungsgemeinschaft within the RTG 2717 program.

\clearpage
\newpage

\bibliography{ref}%

\end{document}